\documentclass[aps,prd,preprint,superscriptaddress,floatfix]{revtex4-1}
\usepackage[english]{babel}
\usepackage[utf8]{inputenc}
\usepackage[T1]{fontenc}
\usepackage{amsmath,amssymb,braket,slashed,mathrsfs}
\usepackage{pifont}

\usepackage[squaren,Gray,cdot,binary]{SIunits}
\usepackage{graphicx}
\graphicspath{{./figures/}}
\usepackage{color,hyperref}
\usepackage{cleveref}

\crefformat{section}{Sec.~#2#1#3}
\crefmultiformat{section}{Sec.~#2#1#3}
    { and~#2#1#3}{, #2#1#3}{ and~#2#1#3}
\crefrangeformat{section}{Sec.~#3#1#4\,--\,#5#2#6}
\crefformat{subsection}{Sec.~#2#1#3}
\crefmultiformat{subsection}{Sec.~#2#1#3}
    { and~#2#1#3}{, #2#1#3}{ and~#2#1#3}
\crefrangeformat{subsection}{Sec.~#3#1#4\,--\,#5#2#6}
\crefformat{subsubsection}{Sec.~#2#1#3}
\crefmultiformat{subsubsection}{Sec.~#2#1#3}
    { and~#2#1#3}{, #2#1#3}{ and~#2#1#3}
\crefrangeformat{subsubsection}{Sec.~#3#1#4\,--\,#5#2#6}
\crefformat{figure}{Fig.~#2#1#3}
\crefmultiformat{figure}{Figs.~#2#1#3}
    { and~#2#1#3}{, #2#1#3}{ and~#2#1#3}
\crefrangeformat{figure}{Figs.~#3#1#4\,--\,#5#2#6}
\crefformat{table}{Table~#2#1#3}
\crefmultiformat{table}{Tabs.~#2#1#3}
    { and~#2#1#3}{, #2#1#3}{ and~#2#1#3}
\crefrangeformat{table}{Tabs.~#3#1#4\,--\,#5#2#6}
\crefformat{equation}{(#2#1#3)}
\crefmultiformat{equation}{(#2#1#3)}{ and~(#2#1#3)}{, (#2#1#3)}{ and~(#2#1#3)}
\crefrangeformat{equation}{(#3#1#4)--(#5#2#6)}

\definecolor{linkcolor}{rgb}{.17578125,.1875,.5703125}
\hypersetup{
    pdfstartview={FitH},    
    pdftitle={Prospects for a lattice computation of 
              rare kaon decay amplitudes (I)},    
    pdfauthor={RBC and UKQCD collaborations},           
    pdfsubject={Physics article},                
    pdfcreator={Antonin Portelli},          
    pdfkeywords={particle} {physics} {quantum} {chromodynamics} {QCD}  {hadrons} {kaons} {rare decays} {flavor} {FCNC} {lattice}, 
    colorlinks=true,        
    linkcolor=linkcolor,    
    citecolor=linkcolor,    
    filecolor=black,        
    urlcolor=linkcolor
}

\newcommand{\cf}{\textit{cf.}~}

\renewcommand{\bar}{\overline}

\newcommand{\diff}{\mathrm{d}}
\renewcommand{\epsilon}{\varepsilon}

\DeclareMathOperator{\timeor}{T}

\renewcommand{\eqref}[1]{(\ref{#1})}

\newcommand{\cu}{Physics Department, Columbia University, New York, 
                 NY 10027, USA}
\newcommand{\soton}{School of Physics and Astronomy, University of Southampton, 
                    Southampton SO17 1BJ, UK}

\begin{document}
    \title{Prospects for a lattice computation of rare kaon decay amplitudes\\
           $K\to\pi\ell^+\ell^-$ decays}
    \author{N.H.~Christ}\affiliation{\cu}
    \author{X.~Feng}\affiliation{\cu}
    \author{A.~Portelli}\affiliation{\soton}
    \author{C.T.~Sachrajda}\affiliation{\soton}
    \collaboration{RBC and UKQCD collaborations}
    \date{\today\\[0.5in]}
    \pacs{PACS}

    \begin{abstract}
      The rare kaon decays $K\to\pi\ell^+\ell^-$ and $K\to\pi\nu\bar{\nu}$ are
      \emph{flavor changing neutral current} (FCNC) processes and hence
      promising channels with which to probe the limits of the standard model
      and to look for signs of new physics. In this paper we demonstrate the
      feasibility of lattice calculations of $K\to\pi\ell^+\ell^-$ decay
      amplitudes for which long-distance contributions are very significant. We
      show that the dominant finite-volume corrections (those decreasing as
      powers of the volume) are negligibly small and that, in the four-flavor
      theory, no new ultraviolet divergences appear as the electromagnetic
      current $J$ and the effective weak Hamiltonian $H_W$ approach each other.
      In addition, we demonstrate that one can remove the unphysical terms
      which grow exponentially with the range of the integration over the time
      separation between $J$ and $H_W$. We will now proceed to exploratory
      numerical studies with the aim of motivating further experimental
      measurements of these decays. Our work extends the earlier study by
      Isidori, Turchetti and Martinelli~\citep{Isidori:2005tv} which focussed
      largely on the renormalization of ultraviolet divergences. In a companion
      paper~\citep{Christ:2015} we discuss the evaluation of the long-distance
      contributions to $K\to\pi\nu\bar{\nu}$ decays; these contributions are
      expected to be at the level of a few percent for $K^+$ decays.
    \end{abstract}
    \maketitle
    %
    \section{Introduction}
    The rare kaon decays $K\to\pi\ell^+\ell^-$ and $K\to\pi \nu\bar{\nu}$ are
    flavor changing neutral current (FCNC) processes which arise in the
    Standard Model through $W$-$W$ and $\gamma/Z$-exchange diagrams, containing
    up, charm and top quarks in the loop. As a second-order electroweak
    interaction, the SM contributions are highly suppressed in FCNC processes,
    leaving the rare kaon decays as ideal probes for the observation of New
    Physics effects. Additionally, these decays can be used to determine SM
    parameters such as $V_{td}$ and $V_{ts}$, to search for CP violating
    effects and to test the low-energy structure of QCD as described within the
    framework of chiral perturbation theory (ChPT). In this paper we discuss
    the feasibility of computing $K\to\pi\ell^+\ell^-$ decay amplitudes in
    lattice simulations; the corresponding study for $K\to\pi \nu\bar{\nu}$
    decays will be presented in a forthcoming companion
    paper~\cite{Christ:2015}.

    The first observation of 41 $K^+\to\pi^+e^+e^-$ decays was made at the CERN
    PS accelerator in 1975~\cite{Bloch:1974ua}. After a long series of
    experiments spanning 40 years, NA48/2 at the CERN SPS accelerator has
    observed a sample of 7253 $K^\pm\to\pi^\pm e^+e^-$
    decays~\cite{Batley:2009aa} and a sample of 3120
    $K^\pm\to\pi^\pm\mu^+\mu^-$ decays~\cite{Batley:2011zz}. These precision
    measurements give important information on the low-energy structure of the
    weak interaction and provide important tests of ChPT. Though expected to be
    difficult, the first observations of the decays $K_S\to\pi^0e^+e^-$ (7
    events)~\cite{Batley:2003mu} and $K_S\to\pi^0\mu^+\mu^-$ (6
    events)~\cite{Batley:2004wg} are reported by the NA48/1 experiment at the
    CERN SPS accelerator. These $K_S$ decays are important in isolating the
    contribution of direct CP violation in the decay $K_L\to\pi^0\ell^+\ell^-$.
    For these interesting CP-violating $K_L$ decays, upper bounds are set for
    ${\rm Br}(K_L\to\pi^0e^+e^-)<2.8\times10^{-10}$~\cite{AlaviHarati:2003mr}
    and ${\rm
    Br}(K_L\to\pi^0\mu^+\mu^-)<3.8\times10^{-10}$~\cite{AlaviHarati:2000hs}.

    On the theoretical side, much work has been done to understand and evaluate
    both the short- and long-distance contributions to rare kaon decays. Some
    useful reviews can be found in
    ~\cite{Buchalla:1995vs,D'Ambrosio:1996nm,Buchholz:1997da,Buras:1997fb,
    Barker:2000gd,Cirigliano:2011ny}. The CP-conserving decays
    $K^+\to\pi^+\ell^+\ell^-$ and $K_S\to\pi^0\ell^+\ell^-$ are dominated by
    long-distance hadronic effects induced through the one-photon exchange
    amplitude. So far the relevant decay amplitudes are studied and
    parameterized within the framework of
    ChPT~\cite{Dib:1988js,D'Ambrosio:1998yj}. A challenge, but also an
    opportunity, for the lattice QCD community is to compute the decay
    amplitudes reliably, as well as determining the necessary low-energy
    constants used in ChPT. We will explain in the following section that the
    significant interference between the direct and indirect CP violating
    components of the decay $K_L\to\pi^0\ell^+\ell^-$ (see
    Refs.~\cite{Isidori:2004rb,Mescia:2006jd}) implies that lattice QCD results
    for $K_S$ decays can be used to evaluate the CP-violating contributions to
    $K_L$ decays. In Ref.~\cite{Isidori:2005tv} it had already been proposed to
    calculate the long-distance contributions to rare kaon decays using lattice
    QCD. Our work builds on Ref.~\cite{Isidori:2005tv} and leads us to conclude
    that such computations are feasible with present understanding and recent
    theoretical and technical advances.
    
    The remainder of this paper is organized as follows. In the next section we
    review the phenomenological background for $K\to\pi\ell^+\ell^-$ decays
    with an emphasis on the long-distance contributions. The procedure to
    obtain the amplitudes from lattice simulations is presented in
    \cref{sec:methodology}. Finite-volume effects and renormalization are
    briefly discussed in \cref{sec:fve,sec:renormalization} respectively.
    Finally we present our conclusions in \cref{sec:concs}.
    \section{Phenomenological background}\label{sec:phenomenology}
    In $K\to\pi\ell^+\ell^-$ decays, the loop function associated with the
    $\gamma$-exchange diagrams has a logarithmic dependence on the masses of
    the quarks entering in the FCNC process (this is known as logarithmic or
    soft Glashow, Iliopoulos, Maiani (GIM) mechanism). The unsuppressed
    sensitivity to the light-quark mass is a signal of long-distance dominance
    in the CP-conserving $K^+\to\pi^+\ell^+\ell^-$ and
    $K_S\to\pi^0\ell^+\ell^-$ decays. The short-distance contribution to the
    amplitude from $Z$-exchange and $W$-$W$ diagrams also exists, but is much
    smaller than the long-distance part induced by the $\gamma$-exchange
    diagrams and does not play an important role in the total branching ratio.
    This logarithmic GIM mechanism does not apply to direct CP violation in
    $K_L\to\pi^0\ell^+\ell^-$ decays since ${\rm Im}\,\lambda_u=0$ where
    $\lambda_q=V_{qs}^*V_{qd}$. As a result, the direct CP-violating
    contribution is short-distance dominated.

\begin{figure}[t]
        \centering\includegraphics{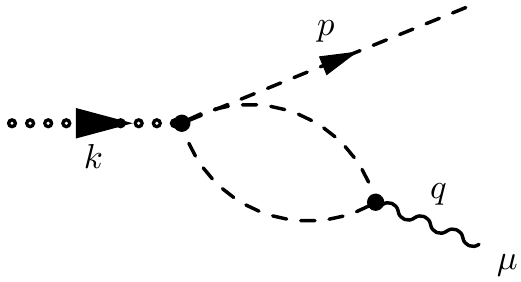}
        \caption{One-loop contribution in ChPT to the
        decays $K_{S}\to\pi^0\ell^+\ell^-$ and $K^{+}\to\pi^+\ell^+\ell^-$. The
        dashed, circled and wiggly lines represent the pions, kaon and photon
        respectively. There is a branch cut when $q^2>4M_\pi^2$.}
        \label{fig:Vpipi}
    \end{figure}

    Considering only the dominant one-photon exchange contribution, the
    amplitude $A_i$ ($i=+,S$) for $K^+$ and $K_S$ decays can be written in
    terms of an electromagnetic transition form factor
    $V(z)$~\cite{Ecker:1987qi,D'Ambrosio:1998yj}
    \begin{equation}   
        A_i=-\frac{G_F\alpha}{4\pi}\,V_i(z)(k+p)^\mu~\bar{u}_{\ell}(p_-)
        \gamma_\mu v_{\ell}(p_+)\,,
    \end{equation}
    where $z=q^2/M_K^2$ and $q=k-p$. Here we follow the notation used in
    Ref.~\cite{Cirigliano:2011ny}; $k$, $p$ and $p_\pm$ indicate the momenta of
    the $K$, $\pi$ and $\ell^\pm$, respectively. For both $K^+$ and $K_S$
    decays, the form factor $V_i(z)$ has been analyzed in
    ChPT~\cite{D'Ambrosio:1998yj} and has been parameterized in the form
    \begin{equation}
      \label{eq:parameterization}
      V_i(z)=a_i+b_iz+V_i^{\pi\pi}(z)\,.
    \end{equation}
    Here $V_i(z)$ is analytic in the complex $z$-plane, with a
    branch cut starting from $4r_\pi^2$, where $r_\pi=M_\pi/M_K$. As shown in
    \cref{fig:Vpipi}, at low energies the $\pi\pi$ intermediate state is
    expected to play the dominate role. Thus $V_i^{\pi\pi}(z)$ is introduced to
    take account of the $\gamma^*\to\pi\pi$ effects. The contribution of
    excited intermediate states is not given explicitly and may be accounted
    for by the assumed polynomial correction
    $a_i+b_iz$.
    A detailed expression for $V_i^{\pi\pi}(z)$ is given
    in~\cite{D'Ambrosio:1998yj}. As a standard twice-subtracted dispersion
    relation, $V_i^{\pi\pi}(z)$ satisfies $V_i^{\pi\pi}(0)=0$. Therefore, the
    inclusion of $V_i^{\pi\pi}(z)$ does not affect $a_i$, which is the form
    factor $V_i(z)$ at zero momentum transfer. The $a_i$ and $b_i$, can be
    determined using the experimental data using the dilepton invariant mass
    spectra as inputs. The parameterization~\cref{eq:parameterization}
    provides a successful description of the $K^+\to\pi^+\ell^+\ell^-$ data but
    shows large corrections beyond leading order in
    ChPT~\cite{D'Ambrosio:1998yj}. A lattice QCD calculation can help to
    understand the origin of these large corrections.

    For $K_S$ decays, only a few events have been observed in experiments. The
    dilepton invariant–mass spectra are therefore unavailable. Assuming vector
    meson dominance, the authors of \cite{D'Ambrosio:1998yj} used the branching
    ratios to determine the parameter $|a_S|$ and obtain
    $|a_S|=1.06^{+0.26}_{-0.21}$ for the electron and
    $|a_S|=1.54^{+0.40}_{-0.32}$ for the muon. There are two drawbacks of using
    the branching ratios: they do not give information about the explicit
    $z$-dependence and only the modulus of $a_S$ can be determined. It would be
    very useful if lattice QCD calculations could determine the sign of $a_S$
    and also provide a test for the vector meson dominance assumption.
    \begin{figure}[t]
        \includegraphics{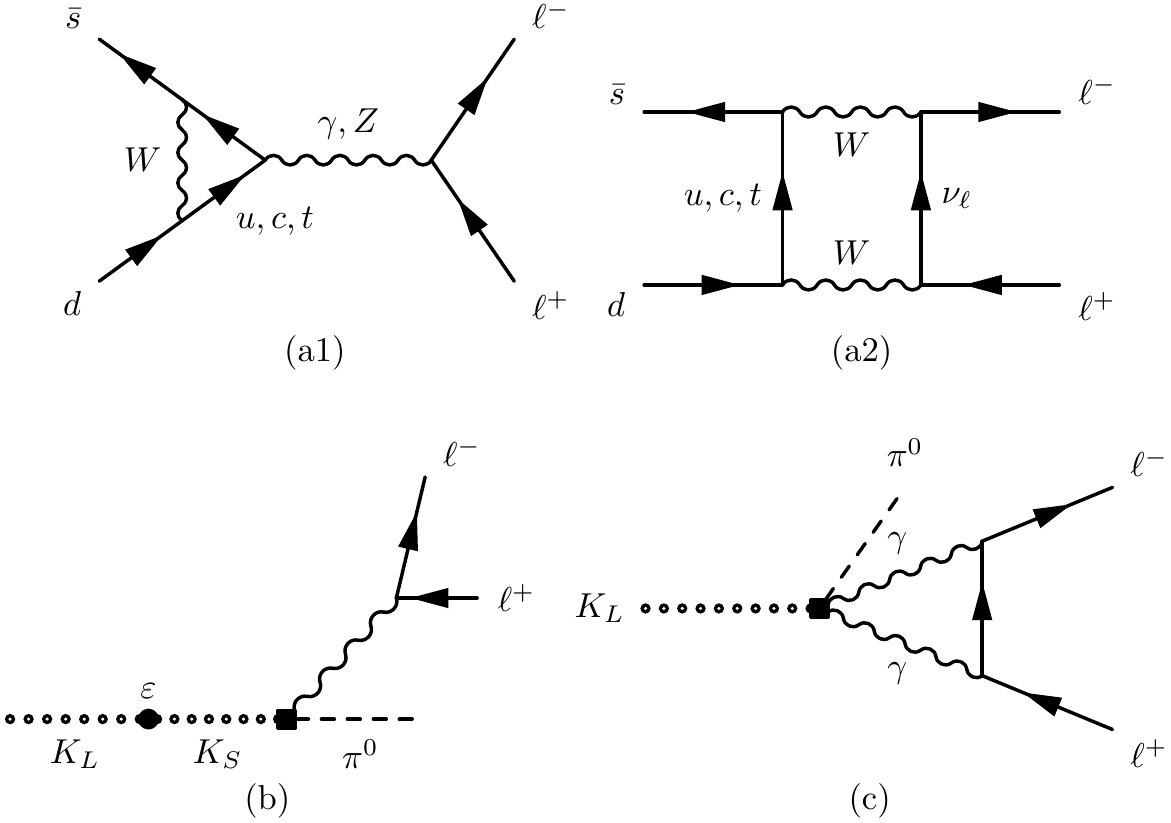}
        \caption{The three major contributions to $K_L\to\pi^0\ell^+\ell^-$
        decays: (a1,a2) short-distance dominated penguin and box diagrams, (b)
        long-distance dominated indirect CP violating contribution occurring
        through $K^0-\bar{K}^0$ mixing, (c) long-distance dominated CP
        conserving component involving two-photon exchange.}
        \label{fig:KL}
    \end{figure}

    $K_L\to\pi^0\ell^+\ell^-$ decays are interesting for precision studies of
    CP violation. The relevant decay amplitudes receive three major
    contributions~\cite{Isidori:2004rb,Mescia:2006jd}: 
    \begin{enumerate}
      \item a short-distance dominated direct CP-violating term, see
      \cref{fig:KL}(a1,a2),
      \item a long-distance dominated indirect CP-violating contribution from
      the decay of the CP-even component of $K_L$ ($K_1\simeq K_S$), see
      \cref{fig:KL}(b),
      \item a CP-conserving component which proceeds through two-photon
      exchanges, see \cref{fig:KL}(c).
    \end{enumerate}
    These three
    contributions are of comparable size~\cite{Isidori:2004rb}. Here we mainly
    focus on the CP violating effects. The total CP-violating contributions to
    the $K_L$ branching ratios are summarized in~\cite{Cirigliano:2011ny} as
    \begin{align} 
      {\rm Br}(K_L\to\pi^0e^+e^-)_{\rm CPV}&=
      10^{-12}\times\left[15.7|a_S|^2\pm6.2|a_S| \left(\frac{{\rm
      Im}\,\lambda_t}{10^{-4}}\right) +2.4\left(\frac{{\rm
      Im}\,\lambda_t}{10^{-4}}\right)^2\right]\\{\rm
      Br}(K_L\to\pi^0\mu^+\mu^-)_{\rm CPV}&=
      10^{-12}\times\left[3.7|a_S|^2\pm1.6|a_S| \left(\frac{{\rm
      Im}\,\lambda_t}{10^{-4}}\right) +1.0\left(\frac{{\rm
      Im}\,\lambda_t}{10^{-4}}\right)^2\right]\,, 
    \end{align} 
    where the $|a_S|^2$-term comes from the indirect CP-violating component of
    the amplitude, the $(\rm Im\,\lambda_t)^2$-term comes from the direct
    CP-violating component and the $|a_S|(\rm Im\,\lambda_t)$-terms are the
    interference between the indirect and direct components. The $\pm$ symbol
    indicates that the sign of $a_S$ is unknown. $|a_S|$ are quantities which
    are expected to be of $O(1)$ and the CKM matrix element takes the value
    ${\rm Im}\,\lambda_t/10^{-4}\approx1.34$. A change in the sign of $a_S$ can
    cause a large difference in the predicted branching ratios for both
    $K_L\to\pi^0 e^+e^-$ and $K_L\to\pi^0\mu^+\mu^-$ decays. Once lattice QCD
    has determined the sign of $a_S$, this large uncertainty will be clarified.
    
    In this paper we do not address the evaluation of the CP-conserving (CPC)
    component of the $K_L\to\pi^0\ell^+\ell^-$ amplitude given by the
    two-photon exchange diagram in \cref{fig:KL}(c).\ The helicity suppression
    of the angular momentum $J=0$ channel leads to a negligible contribution
    (of $O(10^{-14})$ to the branching ratio) for the electron mode, but a
    comparable contribution to the CP-violating one for the muon mode. For
    example, using a phenomenological study based on ChPT, the authors of
    Ref.\,\citep{Isidori:2004rb} estimate \begin{equation} {\rm
    Br}(K_L\to\pi^0\mu^+\mu^-)_{\rm CPC}=\left( 5.2\pm1.6\right)\,10^{-12}\,.
    \end{equation} The $J=2$ contribution on the other hand is expected to be
    negligible for the muon channel, but may be of $O(10^{-13})$ for the
    electron channel so that the decay $K_L\to\pi^0 e^+e^-$ is predominantly
    CP-violating~\citep{Cirigliano:2011ny}. Nevertheless, in due course after
    we manage to compute the $K_L\to\pi^0\gamma^\ast\to\pi^0\ell^+\ell^-$
    contribution to the decay amplitude, it will be an interesting challenge to
    compute the two-photon exchange contribution.
    
    Since the phenomenology of rare kaon decays has, up to now, been based on
    ChPT much of the discussion of this section has been in this context. We
    stress however, that the goal of lattice computations reaches beyond the
    evaluation of the low energy constants. The amplitudes will be calculated
    from first principles at several values of $q^2$ and the results themselves
    can then be used in future phenomenological studies.
    \section{Evaluation of the amplitudes in Euclidean space-time}
    \label{sec:methodology}
    In this section we discuss the evaluation of rare kaon decay amplitudes
    using lattice computations of Euclidean correlation functions. We start
    however, with the definition and a discussion of the amplitudes in
    Minkowski space.
    \subsection{Definition of the amplitude}
    \label{sec:ampdef}
    The long distance part of the $K^+\to\pi^+\ell^+\ell^-$ and
    $K^0\to\pi^0\ell^+\ell^-$ decay amplitudes is given by
    \cite{Ecker:1987qi,D'Ambrosio:1998yj}:
    \begin{equation}
        \label{eq:genamp}
        \mathscr{A}_{\mu}^{j}(q^2)=\int\diff^4x\,
        \bra{\pi^j(\mathbf{p})}\timeor[J_{\mu}(0)H_W(x)]\ket{K^j(\mathbf{k})}\,.
    \end{equation}
    The external states are on their mass shells and we define $q\equiv k-p$.
    The index $j=+,0$ labels the charge of the mesons, and $H_W$ is the
    effective weak Hamiltonian density defined by~\cite{Isidori:2005tv}:
    \begin{equation}
        \label{eq:weakh}
        H_W(x)=\frac{G_F}{\sqrt{2}}\,V_{us}^*V_{ud}\,
        [C_1(Q_1^u-Q_1^c)+C_2(Q_2^u-Q_2^c)]\,,
    \end{equation}
    where the $C_i$ are the Wilson coefficients in a chosen renormalization
    scheme.
    $Q_{1,2}^q$ are the following current-current local operators:
    \begin{equation}\label{eq:Q1Q2def}
        Q_1^q=(\bar{s}_a\gamma_{\mu}^Ld_a)(\bar{q}_b\gamma^{L\,\mu}q_b)
        \quad\quad\text{and}\quad\quad
        Q_2^q=(\bar{s}_a\gamma_{\mu}^Lq_a)(\bar{q}_b\gamma^{L\,\mu}d_b)\,,
    \end{equation}
    where $a$ and $b$ are summed color indices and
    $\gamma_{\mu}^L=\gamma_{\mu}(1-\gamma_5)$. We envisage working in the
    four-flavor theory and exploiting the GIM cancellation of ultraviolet
    divergences as explained in \cref{sec:renormalization}. The electromagnetic
    current $J_{\mu}$ is the standard flavor-diagonal vector current:
    \begin{equation}
        J_{\mu}=\frac{1}{3}(2V_{\mu}^u-V_{\mu}^d+2V_{\mu}^c-V_{\mu}^s)\,.
        \label{eq:Zcurrent}
    \end{equation}
    
    Electromagnetic gauge invariance implies that each non-local matrix element
    can be written in terms of a single invariant form factor:
    \begin{equation}\label{eq:wdef}
      \int\diff^4x\,
        \bra{\pi^j(\mathbf{p})}
        \timeor[J_{\mu}(0)(Q_i^u(x)-Q_i^c(x))]\ket{K^j(\mathbf{k})}\equiv
        \frac{w_i^j(q^2)}{4\pi^2}\left[q^2(k+p)_\mu-(M_K^2-M_\pi^2)q_\mu\right]
    \end{equation}
    and the non-perturbative QCD effects are contained in the form factors
    $w_i^j(q^2)$. In the phenomenological studies described in
    \cref{sec:phenomenology}, the $w_i^j(q^2)$ are written in terms of a
    parametrization influenced by ChPT. Note that a
    consequence of \cref{eq:wdef} is that the matrix elements vanish when
    $\mathbf{p}=\mathbf{k}=\mathbf{0}$, i.e. when the invariant mass of the
    lepton pair is the largest which is kinematically allowed,
    $q^2=q^2_{\textrm{max}}$. Thus to obtain non-zero matrix elements at least
    one of the mesons must have a non-zero three-momentum.
     
    Inserting a complete set of states in each of the two possible
    time-orderings in
    \cref{eq:genamp}, one obtains the following spectral representation for the
    amplitude:
    \begin{align}
        \mathscr{A}_{\mu}^{j}(q^2)&=
        i\int_0^{+\infty}\diff E\,\frac{\rho(E)}{2E}
        \frac{\bra{\pi^j(\mathbf{p})}J_{\mu}(0)\ket{E,\mathbf{k}}
        \bra{E,\mathbf{k}}H_W(0)\ket{K^j(\mathbf{k})}}
        {E_K(\mathbf{k})-E+i\epsilon}\notag\\
        &\quad-i\int_0^{+\infty}\diff E\,\frac{\rho_S(E)}{2E}
        \frac{\bra{\pi^j(\mathbf{p})}H_W(0)\ket{E,\mathbf{p}}
        \bra{E,\mathbf{p}}J_{\mu}(0)\ket{K^j(\mathbf{k})}}
        {E-E_{\pi}(\mathbf{p})+i\epsilon}\,,
        \label{eq:minkspecrep}
    \end{align}
    where $\rho$ and $\rho_S$ are the associated spectral densities. For what
    follows it is important to notice that $\rho$ ($\rho_S$) selects
    only states with strangeness $S=0$ ($S=1$).
    \subsection{Euclidean correlators}\label{ssec:corr}
    For the remainder of the paper we assume that the vector current is the
    conserved one as given by Noether's theorem and which depends on the chosen
    lattice discretization of QCD. The rare kaon decay amplitudes will be
    determined by computing Euclidean correlation functions and we now turn to
    a discussion of these. Throughout this section we consider the time
    dimension to be infinite or very large compared to the time separations of
    the inserted operators.
    \subsubsection{$2$-point functions}
    To obtain the energy of single-meson states we consider the following
    $2$-point function:
    \begin{equation}
        \Gamma^{(2)}_P(t,\mathbf{p})=
        \braket{\phi_P(t,\mathbf{p})\phi^{\dagger}_P(0,\mathbf{p})}\,,
    \end{equation}
    where $\phi_P(t,\mathbf{p})$ is an annihilation operator for a pseudoscalar
    meson $P$ with spatial momentum $\mathbf{p}$ at time $t$. For $t\gg 0$ (and
    $t\ll T/2$, where $T$ is the temporal extent of the lattice),
    $\Gamma^{(2)}_P(t,\mathbf{p})$ has the following behavior:
    \begin{equation}
        \Gamma^{(2)}_P(t,\mathbf{p})=\frac{|Z_P|^2}{2E_P(\mathbf{p})}
        e^{-E_P(\mathbf{p})t}\,,
    \end{equation}
    with $Z_P=\bra{0}\phi_P(0,\mathbf{0})\ket{P(E_P(\mathbf{p}),\mathbf{p})}$
    and $E_P(\mathbf{p})=\sqrt{M_P^2+\mathbf{p}^2}$. Depending on the choice of
    interpolating operator $\phi_P$, $Z_P$ may (and in general will) depend on
    the momentum $\mathbf{p}$ but we do not exhibit this dependence explicitly
    here and in the following.
    \subsubsection{$3$-point functions}
\begin{figure}[!t]
        \includegraphics{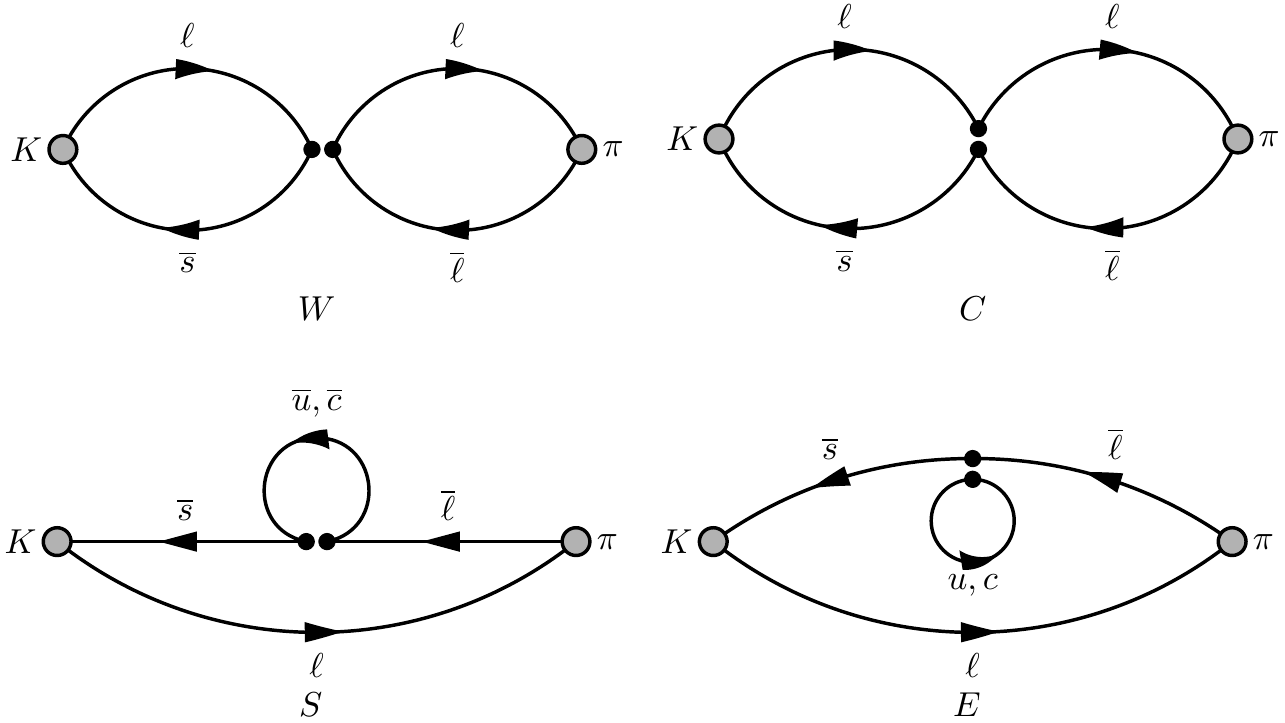}
        \caption{Diagrams contributing to the 3-point function
        $\Gamma^{(3)}_H(t_H,\mathbf{p})$ defined in Eq.\,\cref{eq:3ptHW}. The
        two black circles represent the currents in the four-quark operators
        $Q_{1,2}^q$ defined in \cref{eq:Q1Q2def}. $\ell$ denotes a light
        ($u$ or $d$) quark propagator. The different topologies contain the
        operators $Q_1^q$ or $Q_2^q$ depending on whether the initial state is a
        charged or neutral kaon. For example, when the initial state is $K^+$,
        the $W$ and $S$ topologies contain an insertion of $Q_2^q$ while the
        $C$ and $E$ topologies contain an insertion of $Q_1^q$.}
        \label{fig:3pt_HW_diag}
    \end{figure}

    To obtain matrix elements of the effective weak Hamiltonian, we define
    the following $3$-point function:
    \begin{equation}
        \label{eq:3ptHW}
        \Gamma^{(3)}_H(t_H,\mathbf{p})=\int\diff^3\mathbf{x}\,
                            \braket{\phi_{\pi}(t_{\pi},\mathbf{p})
                    H_W(t_H,\mathbf{x})\phi^{\dagger}_K(0,\mathbf{p})}
    \end{equation}
    with $0<t_H<t_{\pi}$ and on a discrete lattice the integral over
    $\mathbf{x}$ is replaced by the corresponding sum. The $4$ Wick
    contractions necessary to compute $\Gamma^{(3)}_H$ are illustrated in
    \cref{fig:3pt_HW_diag}. For $0\ll t_H\ll t_{\pi}$, $\Gamma^{(3)}_H$ has the
    following behavior:
    \begin{equation}
        \Gamma^{(3)}_H(t_H,\mathbf{p})=\frac{Z_{\pi}Z_K^{\dagger}
        \mathscr{M}_H(\mathbf{p})}{4E_{\pi}(\mathbf{p})E_K(\mathbf{p})}
        e^{-E_{\pi}(\mathbf{p})t_{\pi}}
        e^{-[E_K(\mathbf{p})-E_{\pi}(\mathbf{p})]t_H}\,,
        \label{eq:GHasymp}
    \end{equation}
    with $\mathscr{M}_H(\mathbf{p})=\bra{\pi(\mathbf{p})}H_W(0)
    \ket{K(\mathbf{p})}$.
    We also define the $3$-point function of the electromagnetic
    current:
    \begin{equation}
        \Gamma^{(3)\,P^j}_{J_\mu}(t,t_J,\mathbf{p},\mathbf{k})=
        \int\diff^3\mathbf{x}\,
        e^{-i\mathbf{q}\cdotp\mathbf{x}}
        \braket{\phi_{P^j}(t,\mathbf{p})
        J_{\mu}(t_J,\mathbf{x})\phi^{\dagger}_{P^j}(0,\mathbf{k})}\,,
    \end{equation}
    where $P$ denotes the pseudoscalar meson ($P=\pi$ or $K$) and $j$ its
    charge. This correlation function has the following asymptotic behavior for
    $t\gg t_J\gg 0$:
    \begin{equation}
        \Gamma^{(3)\,P^j}_{J_\mu}(t,t_J,\mathbf{p},\mathbf{k})=
        \frac{|Z_P|^2\mathscr{M}_{J_\mu}^{P^j}(\mathbf{p},\mathbf{k})}
        {4E_{P^j}(\mathbf{p})E_{P^j}(\mathbf{k})}
        e^{-(t-t_J)E_{P^j}(\mathbf{k})}e^{-t_JE_{P^j}(\mathbf{p})}\,,
        \label{eq:GJasymp}
    \end{equation}
    where $\mathscr{M}_{J_\mu}^{P^j}
    (\mathbf{p},\mathbf{k})=\bra{P^j(E_{P^j}(\mathbf{p}),\mathbf{p})}
    J_{\mu}(0)\ket{P^j(E_{P^j}(\mathbf{k}),\mathbf{k})}$ (note that
    $\mathscr{M}_{J_0}^{P^0}(\mathbf{p},\mathbf{p})=0$).
    \subsubsection{$4$-point functions}
    In order to compute the amplitude \cref{eq:genamp}, we define the 
    following unintegrated 4-point correlation function:
    \begin{equation}
        \label{eq:latcorr}
        \Gamma_{\mu}^{(4)\,j}(t_H,t_J,\mathbf{k},\mathbf{p})
        =\int\diff^3\mathbf{x}\int\diff^3\mathbf{y}\,
        e^{-i\mathbf{q}\cdotp\mathbf{x}}
        \braket{\phi_{\pi^j}(t_{\pi},\mathbf{p})
        \timeor[J_{\mu}(t_J,\mathbf{x})H_W(t_H,\mathbf{y})]
        \,\phi_{K^j}^{\dagger}(0,\mathbf{k})}\,,
    \end{equation}
    where $0<t_{J},t_{H}<t_{\pi}$. As explained in the next section,
    the rare kaon decay amplitudes are obtained by integrating
    $\Gamma_{\mu}^{(4)\,j}(t_H,t_J,\mathbf{k},\mathbf{p})$ over $t_H$ and $t_J$
    (or by exploiting time translation symmetry and integrating over their
    difference). 
    
    We now perform the quark Wick contractions in \cref{eq:latcorr} to generate
    the diagrams which need to be evaluated. Assuming isospin symmetry in the
    quark masses, $m_u=m_d$, $20$ types of diagrams have to be computed for the
    charged correlator and $2$ additional ones are needed for the neutral
    correlator. We organize these diagrams in $5$ classes, which are presented
    in \cref{fig:Wdiag,fig:Cdiag,fig:Sdiag,fig:Ediag,fig:pi0diag}. It is
    convenient to define the factor
        \begin{equation}
        Z_{\mathrm{K\pi}}(t_{\pi},\mathbf{k},\mathbf{p})=
        \frac{Z_{\pi}Z_{K}^{\dagger}}{4E_{\pi}(\mathbf{p})E_K(\mathbf{k})}
        e^{-E_{\pi}(\mathbf{p})t_{\pi}}\,,
    \end{equation}
    which represents the propagation of the external pseudoscalar mesons in
    $\Gamma_{\mu}^{(4)\,j}(t_H,t_J,\mathbf{k},\mathbf{p})$. This factor does
    not contribute to the rare kaon decay amplitude and we choose to define the
    normalized unintegrated correlator
    $\tilde{\Gamma}_{\mu}^{(4)\,j}\equiv\Gamma_{\mu}^{(4)\,j}/Z_{K\pi}$. The
    decay amplitudes are obtained by integrating
    $\tilde{\Gamma}_{\mu}^{(4)\,j}$ over $t_H$ and $t_J$ as explained in the
    following subsection. We note however, that if the times are sufficiently
    separated for $\tilde{\Gamma}_{\mu}^{(4)\,j}$ to be dominated by single
    particle intermediate states, then one has:
    \begin{equation}
        \label{eq:GRKasymp}
        \tilde{\Gamma}_{\mu}^{(4)\,j}(t_H,t_J,\mathbf{k},\mathbf{p})=
        \begin{cases}
            \displaystyle
            \frac{\mathscr{M}_H(\mathbf{k})
            \mathscr{M}_{J_\mu}^{\pi^j}(\mathbf{p},\mathbf{k})}
            {2E_{\pi}(\mathbf{k})}
            e^{-E_K(\mathbf{k})t_H}e^{-E_{\pi}(\mathbf{k})(t_J-t_H)}
            e^{E_{\pi}(\mathbf{p})t_J}
            & \text{if}\quad 0\ll t_H\ll t_J\\[1em]
            \displaystyle
            \frac{\mathscr{M}_H(\mathbf{p})
            \mathscr{M}_{J_\mu}^{K^j}(\mathbf{p},\mathbf{k})}
            {2E_{K}(\mathbf{p})}
            e^{-E_K(\mathbf{k})t_J}e^{-E_{K}(\mathbf{p})(t_H-t_J)}
            e^{E_{\pi}(\mathbf{p})t_H}
            & \text{if}\quad t_J\ll t_H \ll t_{\pi}\,.
        \end{cases}
    \end{equation}
    \subsection{Extracting the rare kaon decay amplitude}
    \label{ssec:rkamp}
    In order to obtain the amplitude \cref{eq:genamp}, we need to integrate the
    4-point correlator
    $\tilde{\Gamma}_{\mu}^{(4)\,j}(t_H,t_J,\mathbf{k},\mathbf{p})$, defining
    the integrated correlator by:
    \begin{equation}
        I_{\mu}^{j}(T_a,T_b,\mathbf{k},\mathbf{p})=
        e^{-[E_{\pi}(\mathbf{p})-E_K(\mathbf{k})]t_J}
        \int_{t_J-T_a}^{t_J+T_b}\diff t_H\,
        \tilde{\Gamma}_{\mu}^{(4)\,j}(t_H,t_J,\mathbf{k},\mathbf{p})\,,
        \label{eq:intcorr}
    \end{equation}
    where $T_a,T_b>0$. For $T_a,T_b$ such that $0\ll t_J-T_a<t_J+T_b\ll
    t_{\pi}$, this integrated correlator has the following spectral
    representation:
    \begin{align}
        I_{\mu}^{j}(T_a,T_b,\mathbf{k},\mathbf{p})&=
        -\int_0^{+\infty}\hspace{-10pt}\diff E\,\frac{\rho(E)}{2E}
        \frac{\bra{\pi^j(\mathbf{p})}J_{\mu}(0)\ket{E,\mathbf{k}}
        \bra{E,\mathbf{k}}H_W(0)\ket{K^j(\mathbf{k})}}
        {E_K(\mathbf{k})-E}(1-e^{[E_K(\mathbf{k})-E]T_a})\notag\\
        &\mathbf{\hspace{-0.50in}}+\int_0^{+\infty}\hspace{-10pt}
        \diff E\,\frac{\rho_S(E)}{2E}
        \frac{\bra{\pi^j(\mathbf{p})}H_W(0)\ket{E,\mathbf{p}}
        \bra{E,\mathbf{p}}J_{\mu}(0)\ket{K^j(\mathbf{k})}}
        {E-E_{\pi}(\mathbf{p})}(1-e^{-[E-E_{\pi}(\mathbf{p})]T_b})\,,
        \label{eq:euclidspecrep}
    \end{align}
    where, as in Minkowski space in Eq.\,\cref{eq:minkspecrep}, $\rho$ and
    $\rho_S$ are the spectral densities of non-strange and strangeness $S=1$
    states respectively. In finite-volume, we write the spectral densities as
    $\rho(E)=\sum_n(2E_n)\delta(E-E_n)$ (and similarly for $\rho_S$) so that
    the integrals reduce to sums over the finite-volume states $n$. The
    exponential factor in Eq.\,\cref{eq:intcorr} is introduced to cancel the
    $t_J$ dependence in \cref{eq:euclidspecrep}. To recover the Minkowski
    amplitude \cref{eq:genamp}, one needs to consider the $T_a,T_b\to+\infty$
    limit of $I_{\mu}^{j}(T_a,T_b,\mathbf{k},\mathbf{p})$. Since $\rho$ selects
    states with strangeness $S=0$, the contribution of the states with
    $E<E_K(\mathbf{k})$ diverge exponentially as $T_a\to+\infty$ in the first
    integral of \cref{eq:euclidspecrep}. This is a standard feature in
    the evaluation of long-distance contributions (see
    e.g.\,\cite{Christ:2012se,Bai:2014cva} for a detailed discussion in the
    context of the $K_L$\,-\,$K_S$ mass difference). These contributions must
    be subtracted in order to extract the rare kaon matrix element from
    Euclidean correlation functions. Note that there are no exponentially
    growing terms with $T_b$, since all the strange states with momentum
    $\textbf{p}$ have energies larger than $E_\pi(\textbf{p})$. We define the
    subtracted, integrated correlator
    $\bar{I}_{\mu}^{j}(T_a,T_b,\mathbf{k},\mathbf{p})$ by subtracting all the
    terms which grow exponentially with $T_a$ from the right-hand side of
    \cref{eq:euclidspecrep}. With this definition we can write the rare kaon
    decay amplitude as follows:
    \begin{equation}
         \label{eq:euclidamp}
         \mathscr{A}_{\mu}^{j}(q^2)=-i\lim_{T_{a,b}\to\infty}
         \bar{I}_{\mu}^{j}(T_a,T_b,\mathbf{k},\mathbf{p})\,.
    \end{equation}
    
    How does one compute the subtracted quantity
    $\bar{I}_{\mu}^{j}(T_a,T_b,\mathbf{k},\mathbf{p})$ in practice? For
    physical values of the quark masses, the only intermediate states that can
    generate an exponentially growing term in \cref{eq:euclidspecrep} are
    ones consisting of $1$, $2$ or $3$ pions, the vacuum intermediate state
    being forbidden by parity conservation. We will now discuss each of these
    cases, providing two different approaches for the treatment of the dominant
    single-pion contribution.
    \subsubsection{Removal of the single-pion divergence, first method}
    \label{sssec:pidiv}
    The unphysical divergent term in \cref{eq:euclidspecrep}
    coming from the single pion intermediate state is given by:
    \begin{equation}
        D_{\pi}(T_a,\mathbf{k},\mathbf{p})=
        \frac{1}{2E_{\pi}(\mathbf{k})}
        \frac{\mathscr{M}_{J_\mu}^{\pi^j}(\mathbf{q})
        \mathscr{M}_H(\mathbf{k})}
        {E_K(\mathbf{k})-E_{\pi}(\mathbf{k})}
        e^{[E_K(\mathbf{k})-E_{\pi}(\mathbf{k})]T_a}\,,
        \label{eq:pidiv}
    \end{equation}
    where the notation of \cref{ssec:corr} has been used. Because the pion is
    a stable state in QCD, $D_{\pi}(T_a,\mathbf{k},\mathbf{p})$ can be entirely
    determined by fitting the asymptotic time behavior of the 2 and 3-point
    functions as described in \cref{ssec:corr}.
    \subsubsection{Removal of the single-pion divergence, second method}
    We propose a second method to remove the single-pion divergence which is
    more ``automatic'' than the previous approach. This method is analogous to
    the procedure used in the calculation of the $K_L$\,-\,$K_S$ mass
    difference in \citep{Bai:2014cva}. It is based on an additive shift 
    of the weak Hamiltonian:
    \begin{equation}
        \label{eq:hwshift}
        H_W'(x,\mathbf{k})\equiv H_W(x)+c_s(\mathbf{k})\bar{s}(x)d(x)\,,
    \end{equation}
    where $c_s(\mathbf{k})$ is chosen such that:
    \begin{equation}
        \label{eq:cstune}
        \bra{\pi^j(\mathbf{k})}H_W'(0,\mathbf{k})\ket{K^j(\mathbf{k})}=0\,.
    \end{equation}
   By replacing $H_W$ with $H^\prime_W$ the divergent contribution from the
   single-pion state \cref{eq:pidiv} is cancelled. We now show that the
   transformation \cref{eq:hwshift} does not affect the rare kaon decay
   amplitude. The scalar density appearing in \cref{eq:hwshift} can be written
   as a total divergence using the following vector Ward identity (which is
   satisfied exactly on the lattice):
    \begin{equation}
        i(m_s-m_d)\bar{s}d=\partial_{\mu}V_{\mu}^{\bar{s}d}\,,
    \end{equation}
    where $V_{\mu}^{\bar{s}d}$ is the $\bar{s}d$ flavor non-diagonal vector
    Noether (conserved) current. The relevant matrix elements of the scalar
    density can now be written as :
    \begin{align}
        \bra{\pi^j(\mathbf{p})}\bar{s}(x)d(x)\ket{E,\mathbf{p}}&=
        i\frac{E-E_{\pi}(\mathbf{p})}{m_s-m_d}
        \bra{\pi^j(\mathbf{p})}V_{0}^{\bar{s}d}(x)\ket{E,\mathbf{p}}\\
        \bra{E,\mathbf{k}}\bar{s}(x)d(x)\ket{K^j(\mathbf{k})}&=
        i\frac{E_K(\mathbf{k})-E}{m_s-m_d}
        \bra{E,\mathbf{k}}V_{0}^{\bar{s}d}(x)\ket{K^j(\mathbf{k})}\,.
    \end{align}
    Using \cref{eq:euclidspecrep,eq:euclidamp} we find
    that the total contribution of $c_s(\mathbf{k})\bar{s}d$ to the amplitude
    $\mathscr{A}_{\mu}^{j}(q^2)$ is proportional to:
    \begin{equation}
        \int\diff^3\mathbf{x}\,e^{-i\mathbf{q}\cdotp\mathbf{x}}
        \bra{\pi^j(\mathbf{p})}[J_{\mu}(t_J,\mathbf{x}),
        Q^{\bar{s}d}]\ket{K^j(\mathbf{k})}=0
    \end{equation}
    because of the vanishing commutator between the flavor-diagonal current
    $J_{\mu}$ and the flavor non-diagonal vector charge
    $Q^{\bar{s}d}=\int\,d^3\mathbf{y}\,V_{0}^{\bar{s}d}(t_H,\mathbf{y})$. Thus
    the physical amplitude is invariant under the transformation in
    Eq.\,\cref{eq:hwshift}. This property is independent of the value of
    $c_s(\mathbf{k})$ (and thus from the tuning condition \cref{eq:cstune}).
    \subsubsection{Removal of the two-pion divergence}
    \label{sssec:2pidiv}
\begin{figure}[!t]
        \centering
        \includegraphics{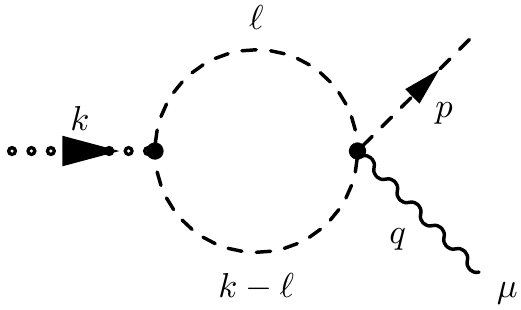}
        \caption{Two-pion intermediate state contribution to the rare kaon
        decay amplitude. The dotted and dashed lines represent respectively the
        kaon and pion propagators.}
        \label{fig:rk_2pi_diag}
    \end{figure}

    In principle, a two pion intermediate state can contribute to a rare kaon
    decay through the process illustrated in \cref{fig:rk_2pi_diag}. The matrix
    elements of vector and axial currents between a single-pion and a two-pion
    state have the following form factor decomposition:
    \begin{equation}
        \bra{\pi(p_1)}V_{\mu}\ket{\pi(p_2)\pi(p_3)}=\epsilon_{\mu\nu\rho\sigma}
        p_1^{\nu}p_2^{\rho}p_3^{\sigma}F(s,t,u)\label{eq:vec2pitopig}
    \end{equation}
    where $s=(p_1+p_2)^2$, $t=(p_1-p_3)^2$ and $u=(p_2-p_3)^2$. 
    
    We now show that the vector current does not contribute.
    Indeed, in \cref{fig:rk_2pi_diag} the $\pi\pi\to\pi\gamma^*$ vertex gives
    the following factor:
    \begin{equation}
        \epsilon_{\mu\nu\rho\sigma}p^{\nu}k^{\rho}
        \int\frac{\diff^4\ell}{(2\pi)^4}
        \frac{\ell^{\sigma}F(s,t,u)}
        {(\ell^2+M_{\pi}^2)[(k-\ell)^2+M_{\pi}^2]}\,.
        \label{eq:vec2pitopigdiag}
    \end{equation}
    Because of $O(4)$ invariance the integral in \cref{eq:vec2pitopigdiag} can
    only be a linear combination of $p^{\sigma}$ and $k^{\sigma}$ which gives a
    vanishing contribution once contracted with the Levi-Civita symbol. 
    
    In the lattice theory, the cubic symmetry is sufficient for the integral
    (or the corresponding sum in a finite volume) to be a vector, but with
    corrections which vanish as the lattice spacing $a\to 0$. At finite lattice
    spacing however, there is a non-zero two-pion contribution from lattice
    artifacts. For example, since the four-component quantity
    $\{(k_1)^3,(k_2)^3,(k_3)^3,(k_4)^3\}$ transforms as the same
    four-dimensional irreducible representation of the cubic group as $k$, one
    can imagine terms of the form $a^2\epsilon_{\mu\nu\rho\sigma}p^{\nu}
    k^{\rho}(k^\sigma)^3$ to be present. These terms will be amplified by the
    growing exponential factor in \cref{eq:euclidspecrep} and will need
    to be considered in the analysis. By studying the behavior with $a^2$ and
    $T_a$ we anticipate being able to confirm our expectation that these
    effects are very small. For example, in our study of $\Delta M_K$, the
    $K_L$-$K_S$ mass difference~\cite{Christ:2012se,Bai:2014cva}, with an
    inverse lattice spacing of 1.73\,GeV and a pion mass of 330\,MeV, we find
    that the on-shell two-pion contributions are just a few per-cent and the
    artifacts are of $O(3\%)$ of these. Assuming similar factors here, the
    exponential factor $e^{[E_K(\mathbf{k})-E]T_a}$ in
    \cref{eq:euclidspecrep} would be insufficient for practical values of
    $T_a$ to make the two-pion contribution significant until the calculations
    reach sub-percent precision.
    \subsubsection{Removal of the three-pion divergence}
    \label{sssec:3pidiv}
\begin{figure}[!t]
        \includegraphics{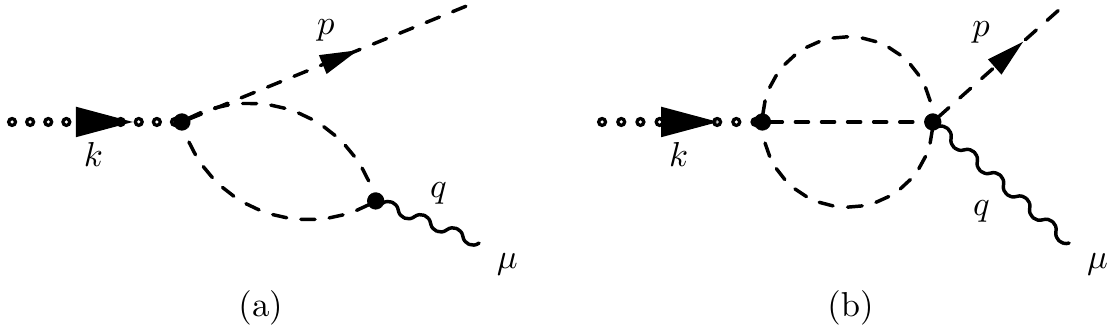}  
        \caption{Examples of contributions from a three-pion
        intermediate state to rare kaon decays.}
        \label{fig:rk_3pi_diag}
    \end{figure}

    Contributions containing three-pion intermediate states are generated by
    diagrams such as those in \cref{fig:rk_3pi_diag}. By comparing the measured
    widths of $K_S\to\pi\pi$ decays to those of $K_{S,+}\to\pi\pi\pi$ decays we
    estimate the relative phase-space suppression to be a factor of $O(1/500)$
    or smaller. Moreover, as explained above, we already expect the on-shell
    two-pion contribution to be very small (of order a few percent) and so we
    anticipate that the on-shell three-pion contribution is negligibly small.
    
    For the the diagram in \cref{fig:rk_3pi_diag}(a) the contribution to the
    growing exponential in \cref{eq:euclidspecrep} can be avoided
    completely by restricting the calculations to $q^2\le4 M_\pi^2$, thus
    cutting out a small region of phase-space. This still allows us to
    determine the amplitudes in most of the $q^2$ range and to compare lattice
    results with ChPT-based phenomenological models and data where this is
    available. Although it is the diagram in \cref{fig:rk_3pi_diag}(a) which is
    dominant in phenomenological analyses based on
    ChPT~\citep{D'Ambrosio:1998yj}, the imaginary part, corresponding to the
    three-pion intermediate state, is neglected.The exponentially growing terms
    from diagrams such as \cref{fig:rk_3pi_diag}(b) cannot be avoided in this
    way and we rely on the phase-space suppression described above. (Moreover,
    much of the exploratory work necessary to develop control of the different
    aspects of the procedure described in this paper, will be performed at
    heavier $u$ and $d$ quark masses, with the mass of the kaon below the
    three-pion threshold.)
    
    In the relatively distant future, if the precision required by experimental
    measurements and achievable in lattice computations is sufficiently high
    one can imagine, in principle at least, reconstructing the contributions
    from three-pion intermediate states explicitly as proposed in
    \cref{sssec:pidiv} and removing the associated exponential divergences in
    \cref{eq:euclidspecrep}. Computing the relevant matrix elements is
    very challenging however, and the recently developed theory of trihadron
    states on a torus \citep{Hansen:2014eka} is, so far at least, purely
    theoretical and has not been used in a practical lattice calculation.
    $K\to\pi\pi\pi$ matrix elements have also not been computed to date.
    \section{Finite volume effects}\label{sec:fve}
    In this section we briefly discuss issues concerning the finite-volume
    corrections arising from the evaluation of rare kaon decay amplitudes
    computed in a finite, periodic hypercubic lattice. The creation of on-shell
    intermediate multi-particle states will generate finite-volume corrections
    which decrease only as powers of the volume and not only exponentially.
    Thus in addition to generating exponentially growing terms in
    \cref{eq:euclidspecrep} (see the discussion in \cref{ssec:rkamp}),
    power corrections are present if there are multi-hadron intermediate states
    with a smaller energy than $M_K$. We therefore have to identify the
    potential on-shell intermediate multi-particle states which can be created
    in rare kaons decays. This is similar to what was done in \cref{ssec:rkamp}
    for the subtraction of unphysical divergences in the Euclidean amplitude.
    We have already shown in \cref{sssec:2pidiv} that with a vector current
    there are no two-pion intermediate states and that for $q^2<4M_\pi^2$ there
    is no power correction from diagrams such \cref{fig:rk_3pi_diag}(a). The
    arguments given in the preceding section that the remaining contributions
    from three-pion intermediate states are negligibly small applies here as
    well. Thus, at the levels of precision likely to be achievable in the near
    future, we do not have to correct for power-like finite-volume effects.
    
    There has been considerable work recently devoted to understanding the
    finite-volume corrections in three-hadron intermediate
    states~\citep{Hansen:2014eka}. If and when the precision of lattice
    computations of rare kaon decays amplitudes reaches the precision requiring
    the control of these effects, then it is to be hoped that the theoretical
    understanding provided in~\citep{Hansen:2014eka} can be developed into a
    practical technique generalizing the use of the Lellouch-L\"uscher factor
    in $K\to\pi\pi$ decay amplitudes~\citep{Lellouch:2000pv}.
    \section{Renormalization}
    \label{sec:renormalization}
    The ultraviolet divergences which appear in the evaluation of the matrix
    elements of the form
     $\int\diff^4x\,\bra{f}\timeor[O_1(0)O_2(x)]\ket{i}$, where
    $O_{1,2}$ are local composite operators and $\ket{i}$, $\ket{f}$
    represent the initial and final states, may come from two sources. Firstly
    $O_{1,2}$ themselves generally require renormalization and secondly
    additional divergences may appear as the two operators approach each other
    in the integral, i.e. as $x\to0$. This is a general feature in the
    evaluation of long-distance contributions to physical processes. In the
    evaluation of the rare kaon decay amplitude \cref{eq:genamp} $O_1$ is the
    vector current and $O_2$ is the effective $\Delta S=1$ Hamiltonian density
    and we start by briefly recalling the normalization of these operators
    before considering the contact terms arising as they approach each other.
    
    \subsection{Renormalization of $H_W$ and $J_{\mu}$}
    We assumed in \cref{ssec:corr} that the current $J_{\mu}$ is the
    conserved one given by Noether's theorem applied to the chosen QCD action
    being used. It therefore satisfies the following vector Ward
    identity:
    \begin{align}
        \partial_{\mu}\braket{V_{\mu}^q(x)\,O(x_1,\dots,x_n)}=0
    \end{align}
    where $O(x_1,\dots,x_n)$ is a multi-local operator with all the points
    $x_1,\dots,x_n$ distinct from $x$. On the lattice, this identity is exactly
    satisfied and the derivative becomes a backward finite difference operator.
    This exact conservation means that the vector current does not require any
    renormalization as the continuum limit is taken.
    
    The Wilson coefficients $C_{1,2}$ in the weak Hamiltonian $H_W$ defined in
    \cref{eq:weakh} are currently known at NLO in the $\bar{\textrm{MS}}$
    scheme~\citep{Buchalla:1995vs}. Since renormalization conditions based
    directly on dimensional regularization, such as the $\bar{\textrm{MS}}$
    scheme, are purely perturbative we envisage following the standard practice
    of renormalizing the bare lattice operators $Q_1^u-Q_1^c$ and $Q_2^u-Q_2^c$
    non-perturbatively into a scheme such as RI-MOM or
    RI-SMOM~\citep{Martinelli:1994ty,Sturm:2009kb,Aoki:2010pe} and then to use
    continuum perturbation theory to match these renormalized operators into
    the $\bar{\textrm{MS}}$ scheme.
   
    The use of a lattice formulation with good chiral symmetry, such as domain
    wall fermions, prevents mixing with dimension-6 operators which transform
    under different representations of the chiral flavor group $SU(4)_L\times
    SU(4)_R$. Within this formulation we can follow the renormalization
    procedure described in detail in~\citep{Christ:2012se,Bai:2014cva} in the
    evaluation of the $K_L$-$K_S$ mass difference. In that case the effective
    weak Hamiltonian is a simple extension of \cref{eq:weakh},
    \begin{equation}\label{eq:weakhp}
      H_W^{\Delta M_K}=
      \frac{G_F}{\sqrt{2}}
      \sum_{q,q^\prime=u,c}V_{qd}V^\ast_{q^\prime s}
      \left(C_1Q_1^{qq^\prime}+C_2Q_2^{qq^\prime}\right)\,
    \end{equation}
    where the operators are generalizations of those in Eq.\,\cref{eq:Q1Q2def} 
    \begin{equation}
      \label{eq:Q1Q2pdef}
        Q_1^{qq^\prime}=
        (\bar{s}_i\gamma_{\mu}^Ld_i)(\bar{q}_j\gamma^{L\,\mu}q^\prime_j)
        \quad\quad\text{and}\quad\quad
        Q_2^{qq^\prime}=
        (\bar{s}_i\gamma_{\mu}^Ld_j)(\bar{q}_j\gamma^{L\,\mu}q^\prime_i)\,.
    \end{equation}
    Since the components with $q\neq q^\prime$ do not contribute to the matrix
    elements for $K\to\pi\ell^+\ell^-$ decays, one is able to rewrite $H_W$ in
    Eq.\,\cref{eq:weakhp} in the form given in Eq.\,\cref{eq:weakh}.
    \subsection{Additional divergences as $H_W(x)$ approaches $J(0)$}
    \begin{figure}[t]
        \centering
            \includegraphics[scale=1]{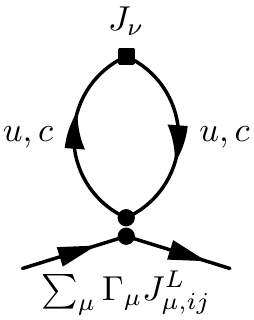}
        \caption{A potentially quadratically divergent insertion into the $S$
        and $E$ classes of diagram. $J_\nu$ represents the conserved
        electromagnetic current and $J^L_{\mu,ij}$ is the local vector current
        $\bar{u}_j\gamma_\mu u_i$ or $\bar{c}_j\gamma_\mu c_i$ from $Q_{1,2}$.}
        \label{fig:ghvp}
    \end{figure}
    In diagrams of the ``loop'' class in topologies $S$ and $E$
    (\cf\cref{fig:Sdiag,fig:Ediag}), there are insertions of the form
    illustrated in \cref{fig:ghvp}. This has been studied in some detail
    in~\citep{Isidori:2005tv} and we briefly summarise the conclusions. The
    vector current $J_\nu$ to which the photon couples is the conserved one
    whereas the vector current $J_\mu^L$ from the weak Hamiltonian is a local
    one; the label $L$ represents \emph{Local}. By power counting the loop
    integral appears to be quadratically divergent. This is reminiscent of the
    evaluation of the one-loop contribution to the vacuum polarization in QED
    and QCD and just as in those cases, electromagnetic gauge invariance
    implies that there is a transversality factor of $q^\mu
    q^\nu-q^2g^{\mu\nu}$ and the order of divergence is reduced by two to a
    logarithmic one. (In momentum space with a lattice action the Ward identity
    $q^\nu J_\nu=0$ becomes $\hat{q}^\nu J_\nu=0$, with $\hat{q}^\nu\equiv
    (2/a)\sin(aq^\nu/2)$.)
    This structure was verified and the divergence explicitly calculated in
    \citep{Isidori:2005tv} in one-loop lattice perturbation theory for Wilson,
    clover and twisted-mass fermions. The logarithmic divergence is mass
    independent, and so cancels exactly in the GIM subtraction between the
    diagrams with $u$ and $c$-quark loops.

    The above argument can be extended straightforwardly to higher-order
    diagrams in which the gluons are contained within the $u$ or $c$ quark loop
    in \cref{fig:ghvp}. The emission of one or more gluons from the $u$ or $c$
    propagators in the loop to be absorbed by a quark or gluon propagator which
    is external to the loop reduces the order of divergence, again leading to a
    convergent loop integration as $J_\nu(x)$ approaches $H_W$. The remaining
    divergences are those which are associated with the renormalization of
    $H_W$.

    We have seen that as a result of gauge invariance and the GIM mechanism in
    the four-flavor theory there are no additional UV divergences in $\int
    \diff^4x\,\bra{\pi}T[J(0)\,H_W(x)]\ket{K}$ coming from the short distance
    region $x\simeq 0$. In the three-flavor theory, gauge invariance still
    protects the correlation function from quadratic divergences, but then
    there remains a logarithmic term which can be removed using
    non-perturbative renormalization techniques~\citep{Martinelli:1994ty}.
    \section{Conclusions}\label{sec:concs}
    Precision flavor physics will continue to be a central tool in searches for
    \emph{new physics} and in guiding and constraining the construction of
    theories beyond the Standard Model. Lattice QCD simulations play an
    important r\^ole in quantifying the non-perturbative hadronic effects in
    weak processes. We must therefore continue to both improve the precision of
    the determination of standard quantities (such as leptonic decay constants,
    semileptonic form factors, neutral meson mixing amplitudes etc.) and to
    extend the range of physical quantities which become amenable to lattice
    simulations. In this paper we propose a procedure for the evaluation of the
    long-distance effects in the rare kaon decay amplitudes
    $K\to\pi\ell^+\ell^-$. These effects represent a significant (and unknown)
    fraction of the amplitudes. In a companion paper \citep{Christ:2015} we
    discuss the prospects for the evaluation of long distance contributions to
    the rare decays $K\to\pi\nu\bar{\nu}$ which will soon be measured by the
    NA-62 experiment at CERN and the KOTO experiment at J-PARC. These decays
    are dominated by short-distance contributions, but given that they will
    soon be measured, it is interesting also to determine the long-distance
    effects which are expected to be of the order of a few percent for $K^+$
    decays.

    In the previous sections we have explained how the technical issues needed
    to perform the lattice simulations can be resolved. Unphysical terms which
    grow exponentially with the range of the time integration, generally
    present when evaluating long-distance effects containing intermediate
    states with energies which are less than those of the external states, were
    shown in \cref{ssec:rkamp} to be absent or small. They could
    potentially arise from the presence of intermediate states consisting of
    one, two or three pions and we considered each of these cases in turn.
    Similarly, the corresponding finite-volume corrections are small provided
    that the invariant mass of the lepton-pair is smaller than $2M_\pi$.
    Ultraviolet effects were discussed in \cref{sec:renormalization}. We
    envisage using the lattice conserved electromagnetic vector current $J_\mu$
    so no renormalization of this operator is required. In addition to the now
    standard renormalization of the weak Hamiltonian $H_W$, we need to consider
    the possible additional ultraviolet divergences which may arise when
    $J_\mu$ and $H_W$ approach each other. Electromagnetic gauge invariance
    implies that no quadratic divergence is present~\cite{Isidori:2005tv} and
    in the four-flavor theory the remaining logarithmic divergence is cancelled
    by the GIM mechanism.

    We conclude that it is now feasible to begin studying rare kaon decays
    $K\to\pi\ell^+\ell^-$ in lattice simulations and in particular to computing
    the long-distance contributions. The next step is to figure out how to
    practically implement the numerical calculation of these amplitudes. The
    main challenge resides in the calculation of the 4-point function in
    Eq.~\cref{eq:latcorr}. One important problem is the evaluation of diagrams
    containing closed loops which require the knowledge of quark propagators
    from a point to itself for every lattice site. Actual numerical
    calculations have been underway for the past two years \citep{Lawson:2015}
    and we are preparing papers describing exploratory results. Within the next
    five years we would hope that the hadronic effects in these decays would be
    quantified with good precision, thus motivating the extension of the
    experimental studies of $K\to\pi\nu\bar\nu$ decays to include also
    $K\to\pi\ell^+\ell^-$ decays.
    \begin{acknowledgments}
    N.H.C. and X.F. were supported in part by US DOE Grant No.DE-SC0011941;
    A.P. and C.T.S. were supported in part by UK STFC Grant ST/L000296/1.
    \end{acknowledgments}
    \newpage
    \appendix
    \section{Feynman diagrams contributing to the rare kaon decay
    correlator}\nopagebreak
    \begin{figure}[h]
        \includegraphics{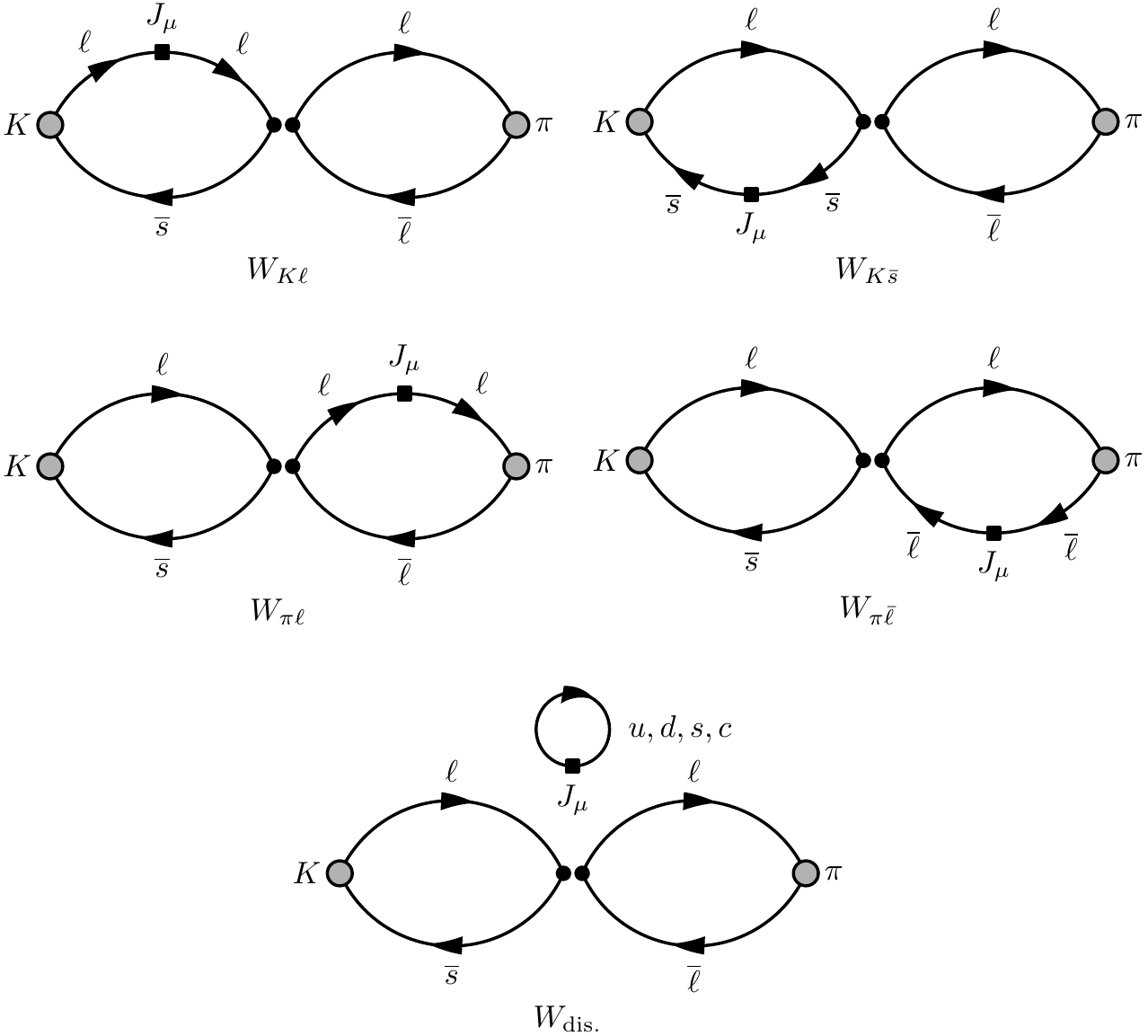}
        \caption{``Wing'' class of diagram contributing to the rare kaon decay
        correlator \cref{eq:latcorr}. 
        The diagrammatic conventions are the same as those in
        \cref{fig:3pt_HW_diag}.}
        \label{fig:Wdiag}
    \end{figure}
    \begin{figure}[p]
        \includegraphics{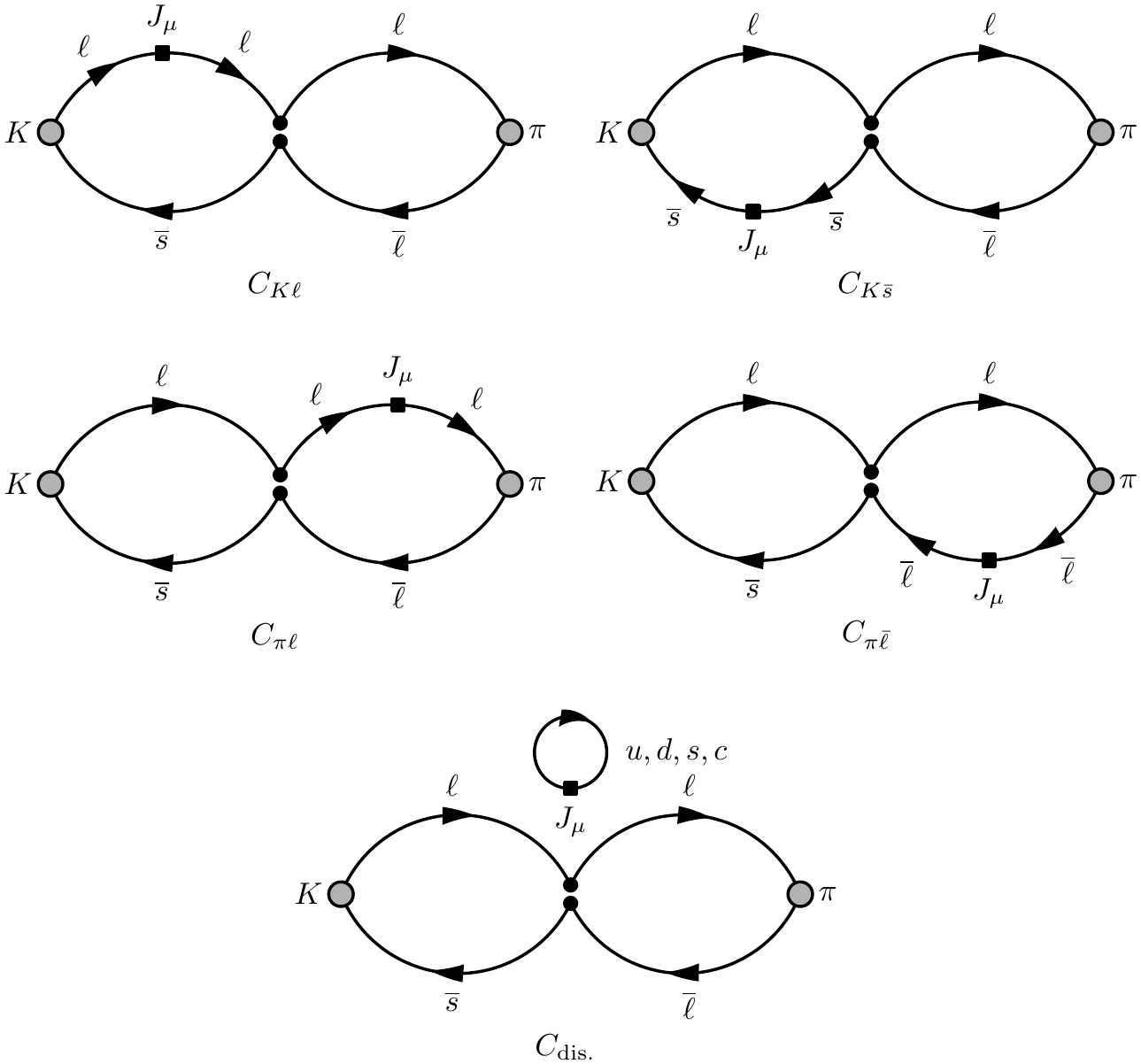}
        \caption{``Connected'' class of diagram contributing to the rare kaon
        decay correlator \cref{eq:latcorr}. The diagrammatic conventions are
        the same as those in \cref{fig:3pt_HW_diag}.The $\ell$ quark is an
        up or down quark depending on the charge of the initial and final
        states.}
        \label{fig:Cdiag}
    \end{figure}
    \begin{figure}[p]
        \includegraphics{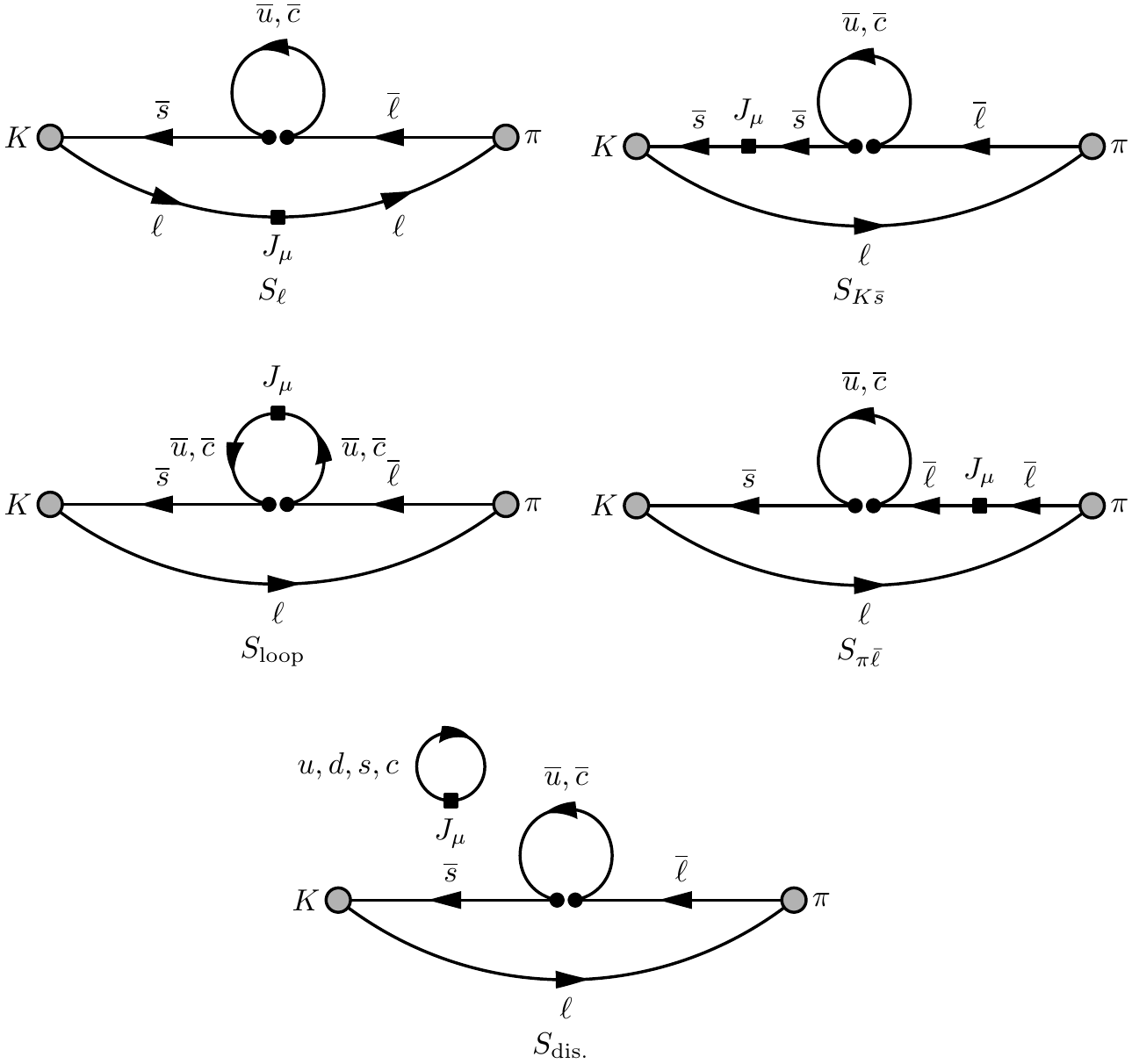}
        \caption{``Saucer'' class of diagram contributing to the rare kaon
        decay correlator \cref{eq:latcorr}. The diagrammatic conventions are
        the same as those in \cref{fig:3pt_HW_diag}.}
        \label{fig:Sdiag}
    \end{figure}
    \begin{figure}[t]
        \includegraphics{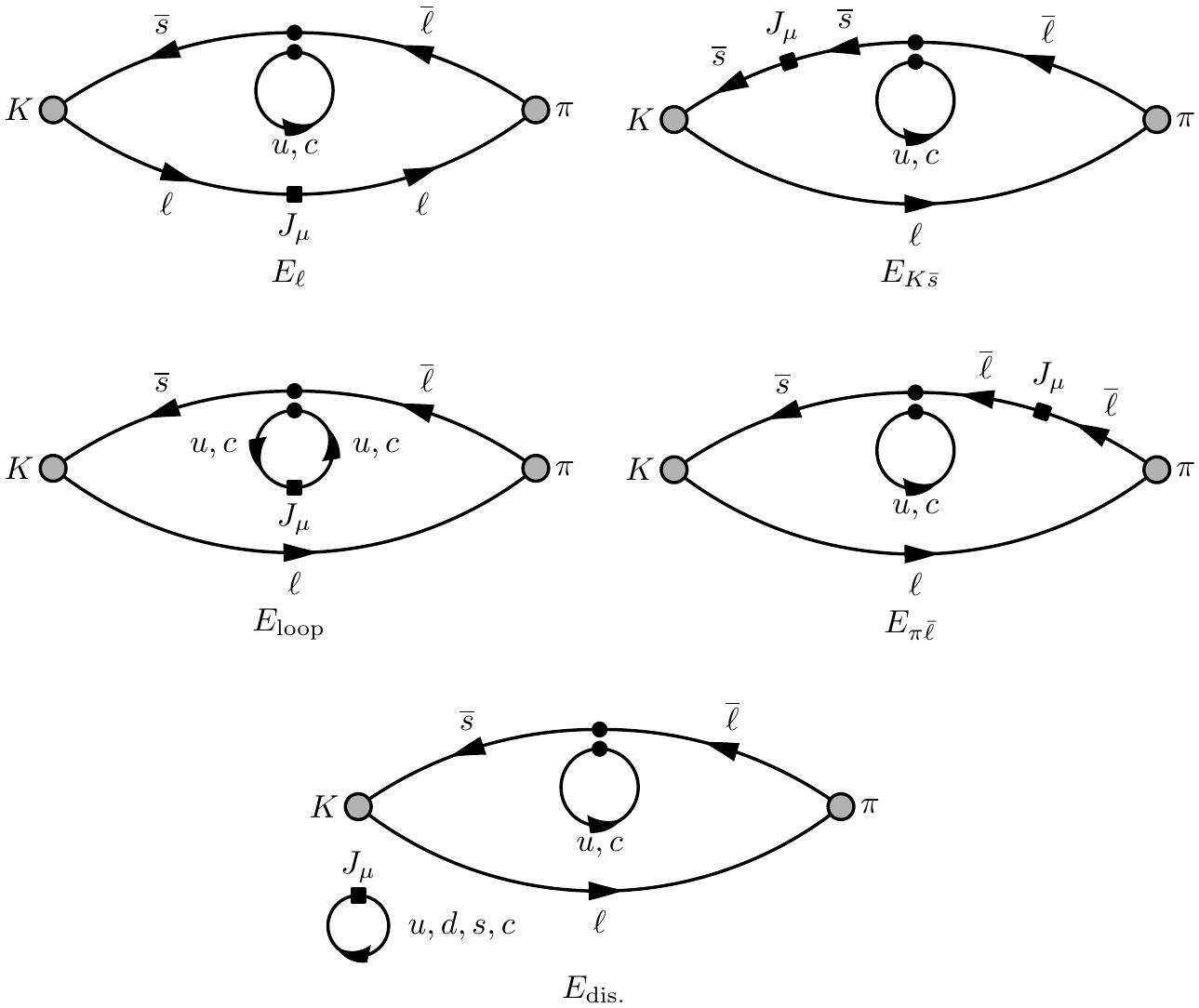}
        \caption{``Eye'' class of diagram contributing to the rare kaon decay
        correlator \cref{eq:latcorr}. The diagrammatic conventions are the same
        as those in \cref{fig:3pt_HW_diag}.}
        \label{fig:Ediag}
    \end{figure}
    \begin{figure}[h]
        \includegraphics{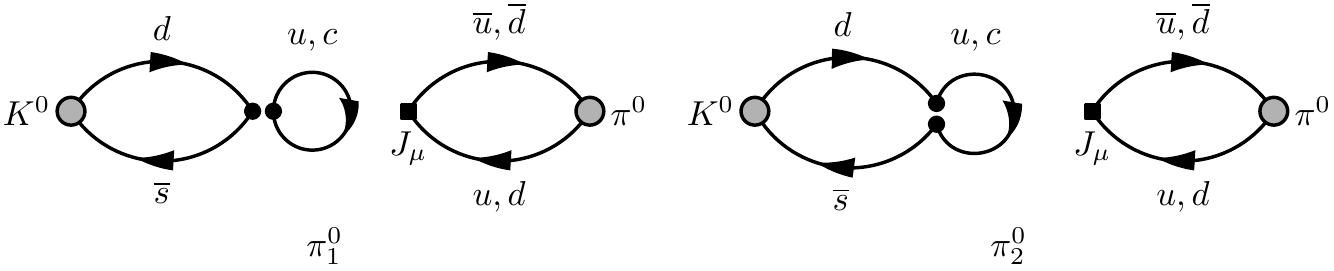}
        \caption{``$\pi^0$\,'' class of diagram contributing to the rare kaon
        decay correlator \cref{eq:latcorr}. These diagrams only contribute to
        the decay of the neutral meson $K^0$. The diagrammatic conventions are
        the same as those in \cref{fig:3pt_HW_diag}.}
        \label{fig:pi0diag}
    \end{figure}
    \newpage
    \bibliography{paper}
\end{document}